\documentclass[preprint,aps,12pt,showpacs,nofootinbib,tightenlines]{revtex4}
\usepackage{amsmath}
\usepackage{amssymb}
\usepackage{CJK}
\usepackage{epsfig}
\usepackage{graphicx}
\usepackage{subfigure}
\usepackage{slashed}
\newcommand{\met}{\not\!\!\!E_{T}}
\begin{document}
\def\pslash{\rlap{\hspace{0.02cm}/}{p}}
\def\eslash{\rlap{\hspace{0.02cm}/}{e}}
\title {Single Higgs boson production at $e^{+}e^{-}$ colliders in the Littlest Higgs Model with T-parity}
\author{Bingfang Yang$^{2,3}$}
\author{Jinzhong Han$^{1}$}\email{hanjinzhongxx@gmail.com}
\author{Sihua Zhou$^{1}$}
\author{Ning Liu$^{3}$}
\affiliation{\footnotesize $^1$ School of Physics and
Electromechnical Engineering, Zhoukou Normal University, Henan, 466001, China\\
$^2$ School of Materials Science and Engineering, Henan
Polytechnic University, Jiaozuo 454000, China\\
$^3$ Institute of Theoretical Physics, Henan Normal University,
Xinxiang 453007, China
   \vspace*{1.5cm}  }

\begin{abstract}

In this work, we investigate the Higgs-boson production processes
$e^{+}e^{-}\rightarrow ZH$, $e^{+}e^{-}\rightarrow
\nu_{e}\bar{\nu_{e}}H$ and $e^{+}e^{-}\rightarrow e^{+}e^{-}H$ in
the littlest Higgs model with T-parity. We present the production
cross sections, the relative corrections and some distributions of
the final states. In the allowed parameter space, we find that the
relative corrections of the three production channels are negative,
the relative correction of the $ZH$ production can reach $-7.5\%$
and the relative corrections of the $\nu_{e}\bar{\nu_{e}}H$ and
$e^{+}e^{-}H$ production can both reach $-6.5\%$ for $\sqrt{s}=500$
GeV with the scale $f=$694 GeV.
\end{abstract}
\pacs{14.80.Ec,12.15.Lk,12.60.-i} \maketitle
\section{ Introduction}
\noindent

In the Standard Model (SM)\cite{sm}, the Higgs mechanism\cite{Higgs
mechanism} leads to the prediction of the Higgs boson. The Higgs
boson is an excitation of the Higgs field, which is an essential
ingredient and will provide direct evidence for the mechanism of
spontaneous symmetry breaking. The SM without the Higgs boson is
incomplete since it predicts massless fermions and gauge bosons.
However, the direct detection of Higgs boson is difficult because it
couples most strongly to the heaviest available channels which will
cascade into complicated multiparticle final states. On the 4th of
July 2012, after a long wait and even generations of immense efforts
by thousands of scientists, CERN announced that both the
ATLAS\cite{ATLAS} and CMS\cite{CMS} experiments had discovered a new
Higgs-like boson, which was a historical event for high-energy
physics.

The Large Hadron Collider(LHC) experiments will determine various
properties of the Higgs boson, up to now, most measurements of this
new particle are consistent with the SM prediction. This corners the
new physics that affects the Higgs couplings to a decoupling region
\cite{higgscoupling}. Due to the clean environment, the complete
profile of the Higgs boson can be precisely studied at an
electron-positron linear collider\cite{higgsee}. In $e^{+}e^{-}$
collider, there are two main production mechanisms for the SM Higgs
boson: Higgs-strahlung and $WW$-fusion. Compared with $WW$-fusion,
the cross section for the similar $ZZ$-fusion process is suppressed
by one order of magnitude. These processes have been studied at
$e^{+}e^{-}$, $e\gamma$ and $\gamma\gamma$ modes in the context of
the SM\cite{zhsm} and the new physics models\cite{zhnp}.

As an extension of the SM, the littlest Higgs model with
T-parity(LHT)\cite{LHT} can successfully solve the electroweak
hierarchy problem and so far remains a popular candidate of new
physics. In the LHT model, some new particles are predicted and some
SM couplings are modified so that the Higgs properties may deviate
from the SM Higgs boson. So the Higgs-boson production processes are
ideal ways to probe the LHT model at the high energy colliders.
These production processes in the LHT model have been studied at the
LHC\cite{HiggsLHC}, but have not been calcualted at the $e^{+}e^{-}$
colliders. In this work, we will study the single Higgs production
processes, $e^{+}e^{-}\rightarrow ZH$, $e^{+}e^{-}\rightarrow
\nu_{e}\bar{\nu_{e}}H$ and $e^{+}e^{-}\rightarrow e^{+}e^{-}H$, in
the LHT model at the $e^{+}e^{-}$ collider.

The paper is organized as follows. In Sec.II we give a brief review
of the LHT model related to our work. In Sec.III  we study the
effects of the LHT model in the single Higgs boson production and
present some discussions of numerical results. Finally, we give a
short summary in Sec.IV.

\section{ A brief review of the LHT model}
\noindent In this section, we only review the LHT model related to
our calculations. For more details, one can refer to
Refs.\cite{LHTphy}.

The LHT model was based on a non-linear $\sigma$ model describing an
$SU(5)/SO(5)$ symmetry breaking, with the global group $SU(5)$ being
spontaneously broken into $SO(5)$ by a $5\times5$ symmetric tensor
at the scale $f\sim \mathcal O$(TeV).

An $[SU(2)\times U(1)]_{1}\times[SU(2)\times U(1)]_{2}$ subgroup of
the $SU(5)$ is gauged and the gauge fields $W_{i \mu}^a$ and
$B_{i\mu}$ $(a = 1, 2, 3,~ i = 1, 2)$ are introduced. In this model,
the action of T-parity on the gauge fields and scalar sector are
defined as:
\begin{eqnarray}
 W_{1\mu}^a \longleftrightarrow W_{2\mu}^a,~~~~
 B_{1\mu} \longleftrightarrow B_{2\mu},~~~~
\Pi \longrightarrow -\Omega \Pi \Omega,
\end{eqnarray}
where $\Omega = {\rm diag}(1,1,-1,1,1)$. The T-odd and T-even gauge
fields can be obtained as
\begin{eqnarray}
 W_L^a &=& \frac{W_1^a + W_2^a}{\sqrt{2}},~~~
 B_L = \frac{B_1 + B_2}{\sqrt{2}},~~~~~ (\mbox{T-even}), \nonumber \\
 W_H^a &=& \frac{W_1^a - W_2^a}{\sqrt{2}},~~~
 B_H = \frac{B_1 - B_2}{\sqrt{2}},~~~~~ (\mbox{T-odd}).
\end{eqnarray}

The electroweak symmetry breaking $SU(2)_L \times U(1)_Y \to
U(1)_{em}$ takes place via the usual Higgs mechanism. The mass
eigenstates of the gauge fields are given by
\begin{eqnarray}
&& W_L^{\pm} = \frac{W_L^1 \mp i W_L^2}{\sqrt{2}},~~~~
  \left( \begin{array}{c} A_L \\ Z_L \end{array} \right)
 = \left( \begin{array}{rc} \cos\theta_W & \sin\theta_W \\
 -\sin\theta_W & \cos\theta_W \end{array} \right)  \left(
 \begin{array}{c} B_L \\ W_L^3 \end{array} \right),~~~~(\mbox{T-even}), \nonumber \\
&& W_H^{\pm} = \frac{W_H^1 \mp i W_H^2}{\sqrt{2}}, ~~~
  \left( \begin{array}{c} A_H \\ Z_H \end{array} \right)
 = \left( \begin{array}{cr} \cos\theta_H & -\sin\theta_H \\
 \sin\theta_H & \cos\theta_H \end{array} \right)
 \left( \begin{array}{c} B_H \\ W_H^3 \end{array} \right),~~~~(\mbox{T-odd}),
\end{eqnarray}
where $\theta_{W}$ is the usual Weinberg angle and $\theta_H$ is the
mixing angle defined by
\begin{eqnarray}
\sin{\theta_{H}} \simeq \frac{5 g g^{\prime}}{4(5 g^2 - g^{\prime
2})} \frac{v_{SM}^2}{f^2},
\end{eqnarray}
where $v_{SM}\simeq 246$ GeV is the SM Higgs vacuum expectation
value (VEV).

To implement T-parity in the fermion sector, it requires the
introduction of the mirror fermions. For each SM $SU(2)_{L}$
doublet, under the $SU(2)_1\times SU(2)_2$ gauge symmetry, a doublet
under $SU(2)_{1}$ and one under $SU(2)_{2}$ are introduced. The
T-parity even combination is associated with the SM $SU(2)_{L}$
doublet while the T-odd combination is given a $\mathcal O(f)$ mass.

In order to avoid dangerous contributions to the Higgs mass from
one-loop quadratic divergences, the third generation Yukawa sector
must be modified. One must also introduce additional singlets
$t'_{1R}$ and $t'_{2R}$ which transform under T-parity as
\begin{equation}
t'_{1R}\leftrightarrow -t'_{2R}
\end{equation}
so the top sector masses can be generated in the following T-parity
invariant way
\begin{eqnarray}
\mathcal{L}_{top} &=& -\frac{1}{2 \sqrt{2}}\lambda_1 f
\epsilon_{ijk} \epsilon_{xy} \big[ (\bar{Q}_1)_i(\Sigma)_{jx}
(\Sigma)_{ky} - (\bar{Q}_2 \Sigma_0)_i (\tilde{\Sigma})_{jx}
(\tilde{\Sigma})_{ky} \big] u_{3R} \nonumber\\&-& \lambda_2 f
(\bar{t}'_1 t'_{1R} + \bar{t}'_2 t'_{2R})+ h.c.
\end{eqnarray}

For the other quarks, it will not be necessary to modify the Yukawa
Lagrangian as in the top sector since their Yukawa coupling is at
least one order of magnitude smaller. Therefore we do not need to
introduce additional singlets for the remaining up-type quarks and
the Yukawa coupling is accordingly given by
\begin{eqnarray}
\mathcal{L}_{up} = -\frac{1}{2\sqrt{2}}\lambda_u
f\epsilon_{ijk}\epsilon_{xy}\left[(\bar
    Q_1)_i(\Sigma)_{jx}(\Sigma)_{ky} - (\bar Q_2 \Sigma_0)_i (\tilde
    \Sigma)_{jx}(\tilde \Sigma)_{ky}\right] u_R+h.c.
\end{eqnarray}

For the down-type quarks, we can construct the Yukawa interaction to
give them masses in the following way:
\begin{eqnarray}
\mathcal{L}_{down}=\frac{i\lambda_d}{2\sqrt{2}}f
\epsilon_{ij}\epsilon_{xyz}\left[(\bar \Psi_2 )_x (\Sigma)_{iy}
  (\Sigma)_{jz}X-(\bar \Psi_1\Sigma_0)_x (\tilde \Sigma)_{iy}(\tilde
 \Sigma)_{jz}\tilde X\right] d_R + h.c.\
\end{eqnarray}

In our calculations, the $Hb\bar{b},HZZ$ and $HWW$ coupling involved
will be different from the SM coupling, which are given by
\begin{eqnarray}
V_{Hb\bar{b}}&=&-\frac{m_{b}}{v}(1-\frac{1}{6}\frac{v^2}{f^2}), \\
V_{HZ_{\mu}Z_{\nu}}&=&\frac{2m_{Z}^{2}}{v}(1-\frac{1}{6}\frac{v^2}{f^2})g_{\mu\nu},\\
V_{HW_{\mu}W_{\nu}}&=&\frac{2m_{W}^{2}}{v}(1-\frac{1}{6}\frac{v^2}{f^2})g_{\mu\nu},
\end{eqnarray}
where $v=v_{SM}(1+\frac{1}{12}\frac{v_{SM}^2}{f^2})$. Although the
differences occur at the order $\mathcal O(v^{2}/f^{2})$, their
contributions cannot be ignored because they appear at the
lowest-order.
\section{Calculation and Numerical results}

\noindent

In the LHT model, the lowest-order Feynman diagrams of the process
$e^{+}e^{-}\rightarrow ZH$, $e^{+}e^{-}\rightarrow
\nu_{e}\bar{\nu_{e}}H$ and $e^{+}e^{-}\rightarrow e^{+}e^{-}H$ are
shown in Fig.1. We can see that the tree-level Feynman diagrams of
these processes in the LHT model are identical with that in the SM.

\begin{figure}[htbp]
\scalebox{0.45}{\epsfig{file=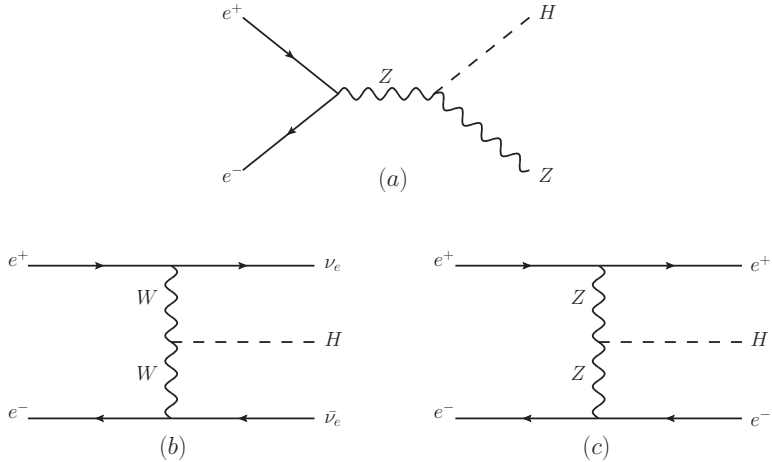}}\vspace{-0.5cm}
\caption{Lowest-order
Feynman diagrams for $e^{+}e^{-}\rightarrow ZH$(a),
$e^{+}e^{-}\rightarrow \nu_{e}\bar{\nu_{e}}H$(b) and
$e^{+}e^{-}\rightarrow e^{+}e^{-}H$(c).}
\end{figure}

In our numerical calculations, the SM parameters are taken as
follows\cite{parameters}
\begin{eqnarray}
\nonumber &&G_{F}=1.16637\times 10^{-5}{\rm GeV}^{-2},
~\sin^{2}\theta_{W}=0.231,~\alpha_{e}=1/128,\\
&&m_{b}=4.65{\rm GeV},~m_{Z}=91.1876{\rm GeV},~m_{H}=126{\rm GeV},\\
&&m_{e}=0.51{\rm MeV},~m_{\mu}=105.66{\rm MeV},~m_{\tau}=1776.82{\rm
MeV}.
\end{eqnarray}

There is only one LHT parameter, the breaking scale $f$, in our
calculation. Considering the constraints in Refs.\cite{constraints},
we choose the relatively relaxed parameter space and vary the scale
in the range $500$ GeV$\leq f\leq 1500$ GeV.

\begin{figure}[htbp]
\begin{center}
\scalebox{0.25}{\epsfig{file=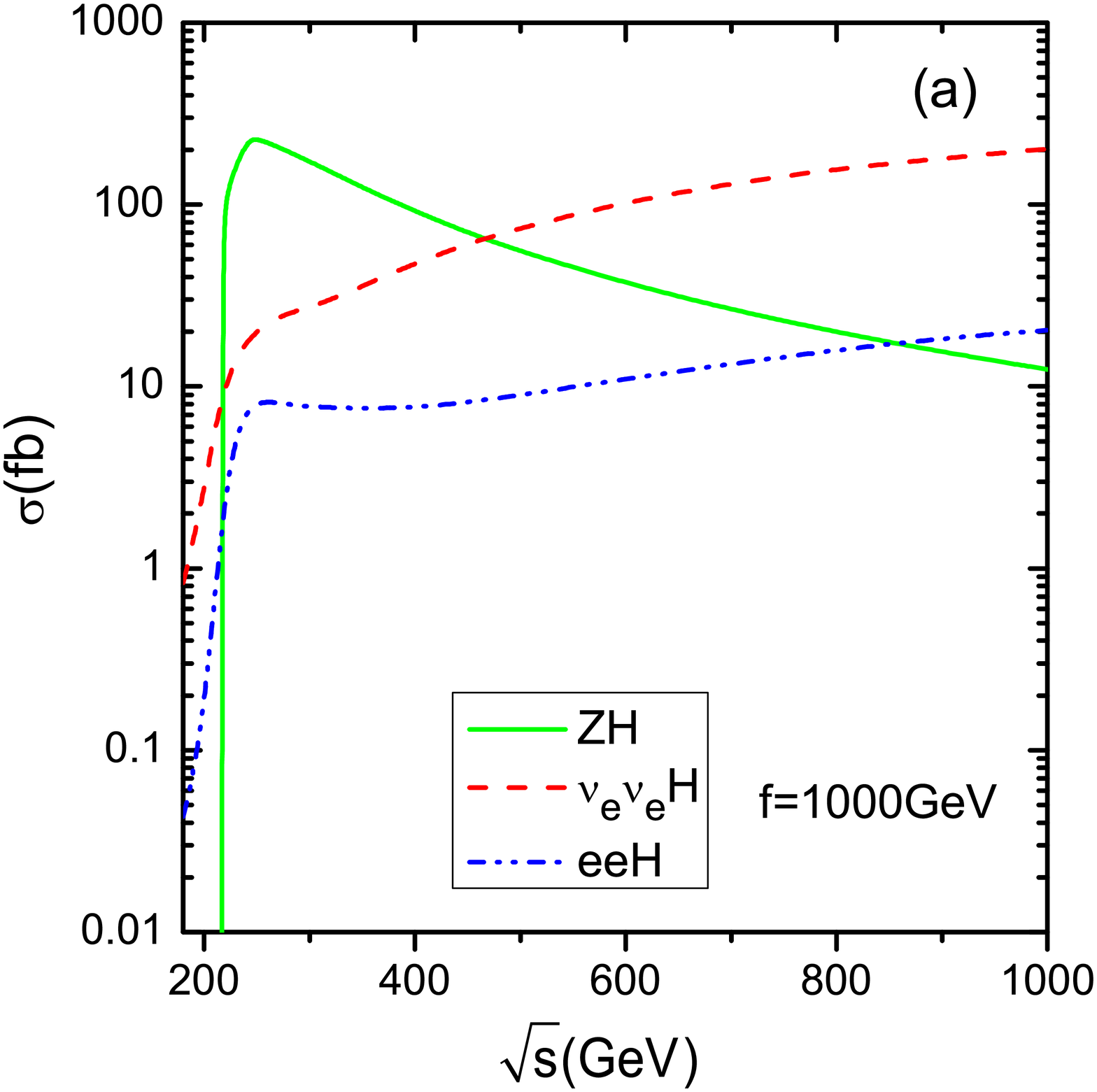}}\vspace{-0.5cm}
\scalebox{0.25}{\epsfig{file=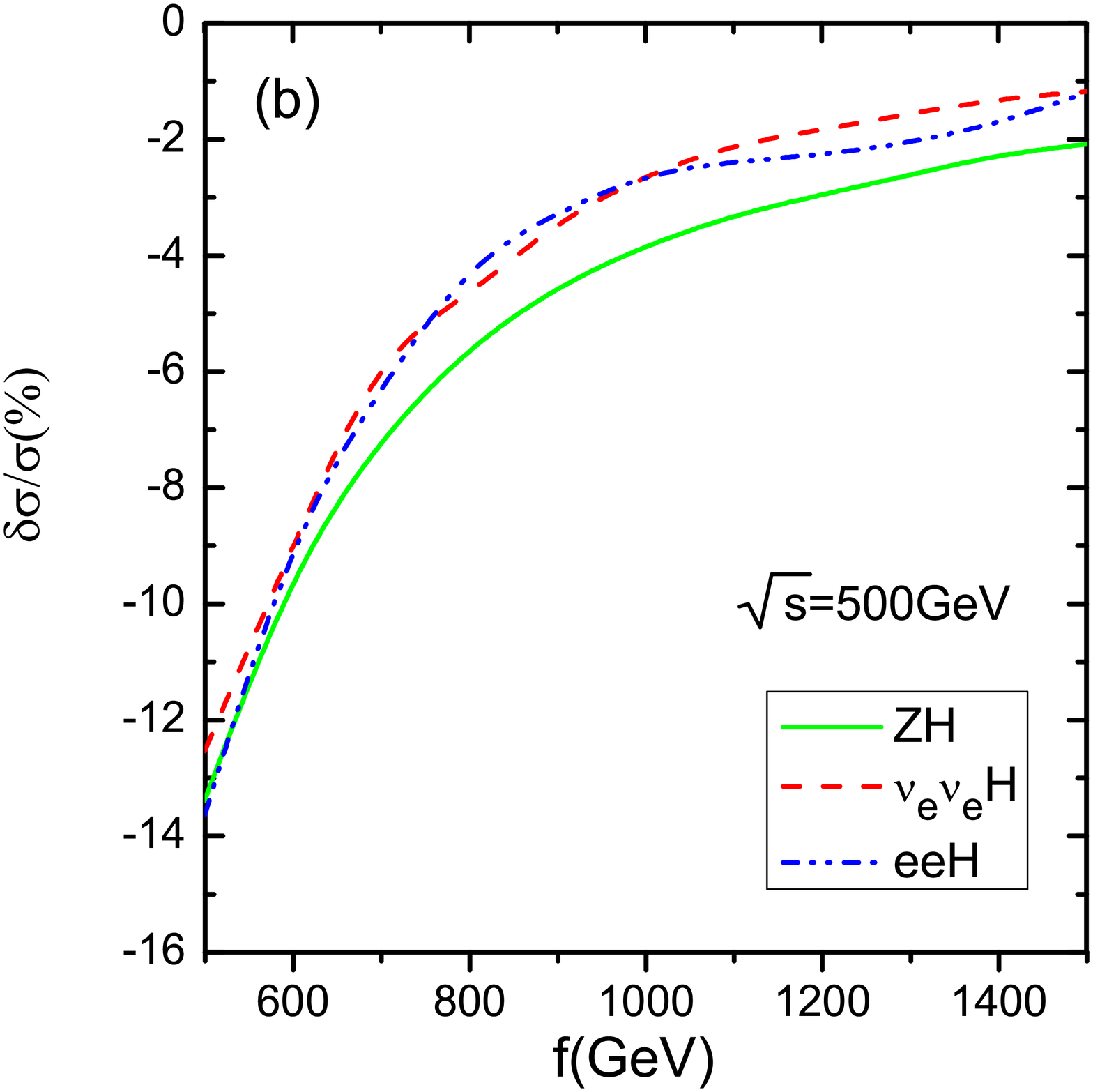}} \caption{The production
cross section $\sigma$ versus the center-of-mass energy $\sqrt{s}$
for $f=1000$ GeV(a) and the relative correction $\delta
\sigma/\sigma$ versus the scale $f$ for $\sqrt{s}=500$ GeV(b) in the
LHT model.}
\end{center}
\end{figure}

In Fig.2(a), we show the dependance of the production cross section
$\sigma$ on the center-of-mass energy $\sqrt{s}$ for the scale
$f=1000$ GeV in the LHT model. We present $ZH$,
$\nu_{e}\bar{\nu}_{e}H$ and $e^{+}e^{-}H$ production channels,
respectively. We can see that the $ZH$ production cross section
dominates at low center-of-mass energies, the corresponding cross
section increases sharply at the threshold and then decreases with
the center-of-mass energy in proportion to $1/s$. The region of
cross section maximum is around $240\sim 250$ GeV and the maximum
value can reach about 230 fb. The $\nu_{e}\bar{\nu_{e}}H$ and
$e^{+}e^{-}H$ production cross section increases with the
center-of-mass energy in proportion to log$(s/m_{H}^{2})$ and hence
becomes more important at energies $\sqrt{s}\geq 500$ GeV.

After the discovery of the Higgs-like boson at the LHC, in order to
study the properties of this new particle with high precision, many
schemes of the so-called Higgs factory have been proposed
\cite{Higgsfactory}. For example, the proposed LEP3 or China Higgs
Factory (CHF) with a center-of-mass energy 240 GeV, the TLEP with a
center-of-mass energy 350 GeV, the ILC with a center-of-mass energy
500 GeV, and so on. In Tab.I, we display the lowest-order dominant
Higgs boson production cross section in the LHT model for different
Higgs factories.

\begin{table}[ht]
\caption{Dominant Higgs-boson production cross section in the LHT
model at various center-of-mass energies of the $e^{+}e^{-}$
collision for $f=1000$ GeV, $m_{H}=126$ GeV. \label{tab1}}
\bigskip
\begin{tabular}{|c|c|c|c|c|}
\hline
  $\sqrt{s}$ [GeV] & $~~~240~~~$ & $~~~350~~~$ & $~~~500~~~$ & $~~~1000~~~$ \\
\hline
  $\sigma(e^{+}e^{-}\rightarrow ZH)$[fb] & 227 & 124 & 55.3 & 12.4\\
\hline
  $\sigma(e^{+}e^{-}\rightarrow \nu_{e}\bar{\nu_{e}} H)$[fb] & 21.1 & 35.7 & 74.6 & 203\\
\hline
  $\sigma(e^{+}e^{-}\rightarrow e^{+}e^{-} H)$[fb] & 7.9 & 7.5 & 8.9 & 20.4\\
\hline
\end{tabular}
\end{table}

In Fig.2(b), we show the dependance of the relative correction
$\delta \sigma/\sigma$ on the scale $f$ for the center-of-mass
energy $\sqrt{s}=500$ GeV. We present the relative correction
$\delta \sigma/\sigma$ of $ZH$, $\nu_{e}\bar{\nu_{e}}H$ and
$e^{+}e^{-}H$ production channels, respectively. We can see that the
relative correction $\delta \sigma/\sigma$ decreases with the scale
$f$ increasing, which means that the correction of the LHT model
decouples with the scale $f$ increasing. For the three production
channels, the relative corrections $\delta \sigma/\sigma$ are all
negative and each of them can maximally reach $-13\%$ when the scale
$f=500$ GeV. Moreover, we can see that the behaviors of the three
production channels are similar due to the similar LHT correction to
the $HZZ$ and $HWW$ couplings.

By combining the Higgs data from the LHC and electroweak precise
measurements, the authors in Ref.\cite{LHTstatus} get the constraint
on the scale $f$ in the LHT model, that is, $f\geq 694$ GeV at
$95\%$ CL. The direct searches for the new particles can also
provide constraints on the scale $f$, but these bounds can be weaken
by the small $k$. So, if we require the scale $f\geq 694$ GeV, the
relative correction $\delta \sigma/\sigma$ of $ZH$ production can
maximally reach $-7.5\%$ and the relative corrections $\delta
\sigma/\sigma$ of $\nu_{e}\bar{\nu_{e}}H$ and $e^{+}e^{-}H$
production can both maximally reach $-6.5\%$.

By exploiting the $HZ\rightarrow Xl^{+}l^{-}$ channel, the
$e^{+}e^{-}\rightarrow ZH$ production cross sections at
$\sqrt{s}=350$ GeV with an integrated luminosity of 500 fb$^{-1}$
can be measured with statistical errors of $2.6\sim 3.1\%$ for
Higgs-boson masses from 120 to 160 GeV\cite{zll}. If the
center-of-mass energy is upgraded to 500 GeV at a linear
$e^{+}e^{-}$ collider, this will allow to measure the Higgs
production cross sections at the level of a few
percent\cite{higgsMeasurement}. So, the LHT effects can be tested at
the future $e^{+}e^{-}$ colliders with a high luminosity.

\begin{figure}[htbp]
\scalebox{0.5}{\epsfig{file=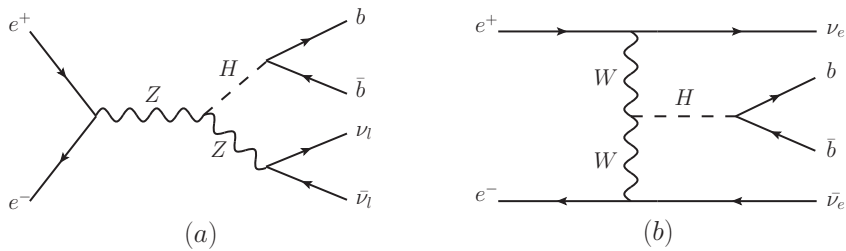}}\vspace{-0.5cm}\caption{
Feynman diagrams for $e^{+}e^{-}\rightarrow ZH$ followed by the
subsequent $Z\rightarrow \nu_{l}\bar{\nu_{l}}, H\rightarrow
b\bar{b}$(a) and $e^{+}e^{-}\rightarrow \nu_{e}\bar{\nu_{e}}H$
followed by the subsequent $ H\rightarrow b\bar{b}$(b).}
\end{figure}

\begin{figure}[htbp]
\begin{center}
\scalebox{0.25}{\epsfig{file=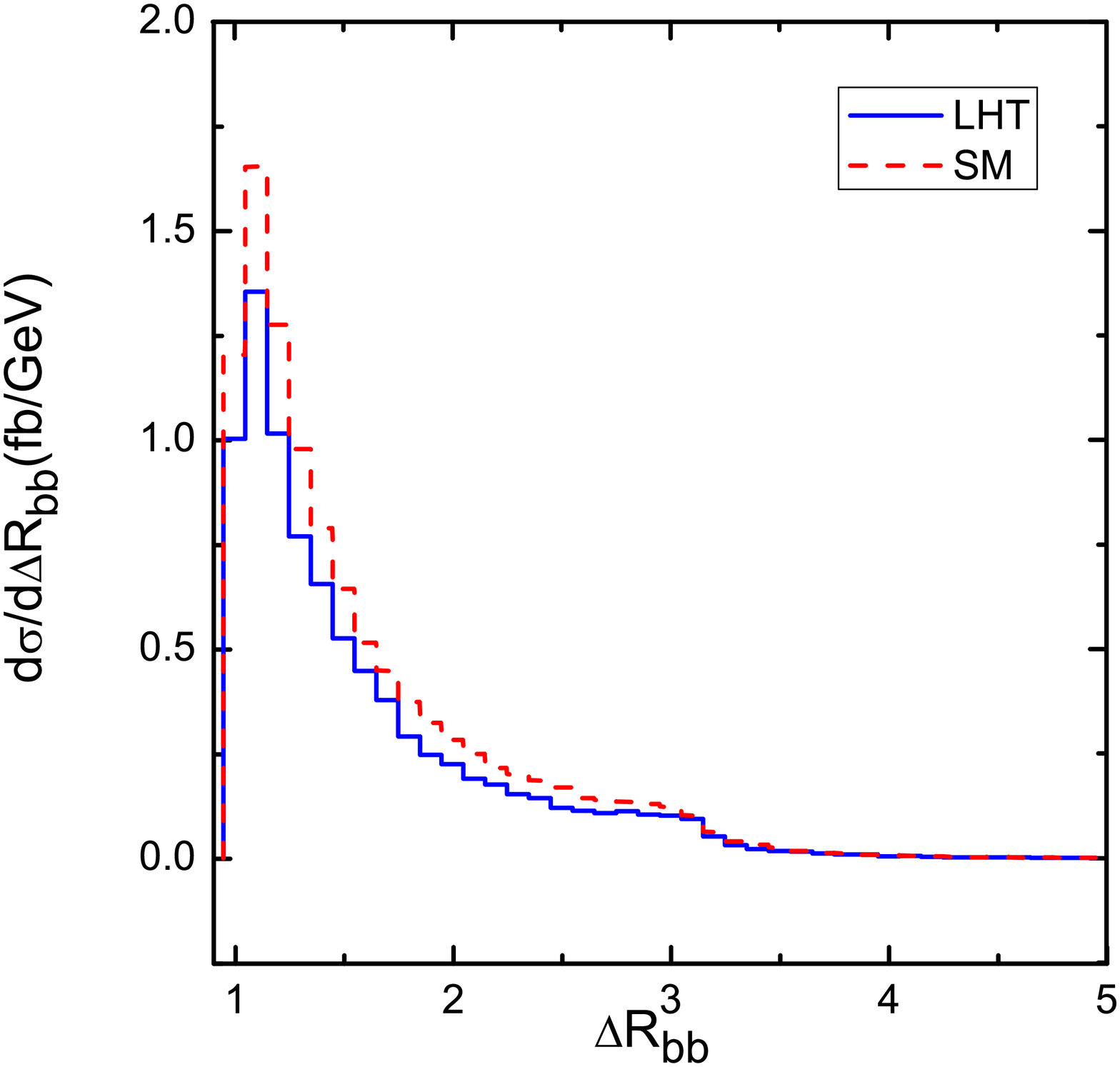}}\hspace{-0.8cm}
\scalebox{0.25}{\epsfig{file=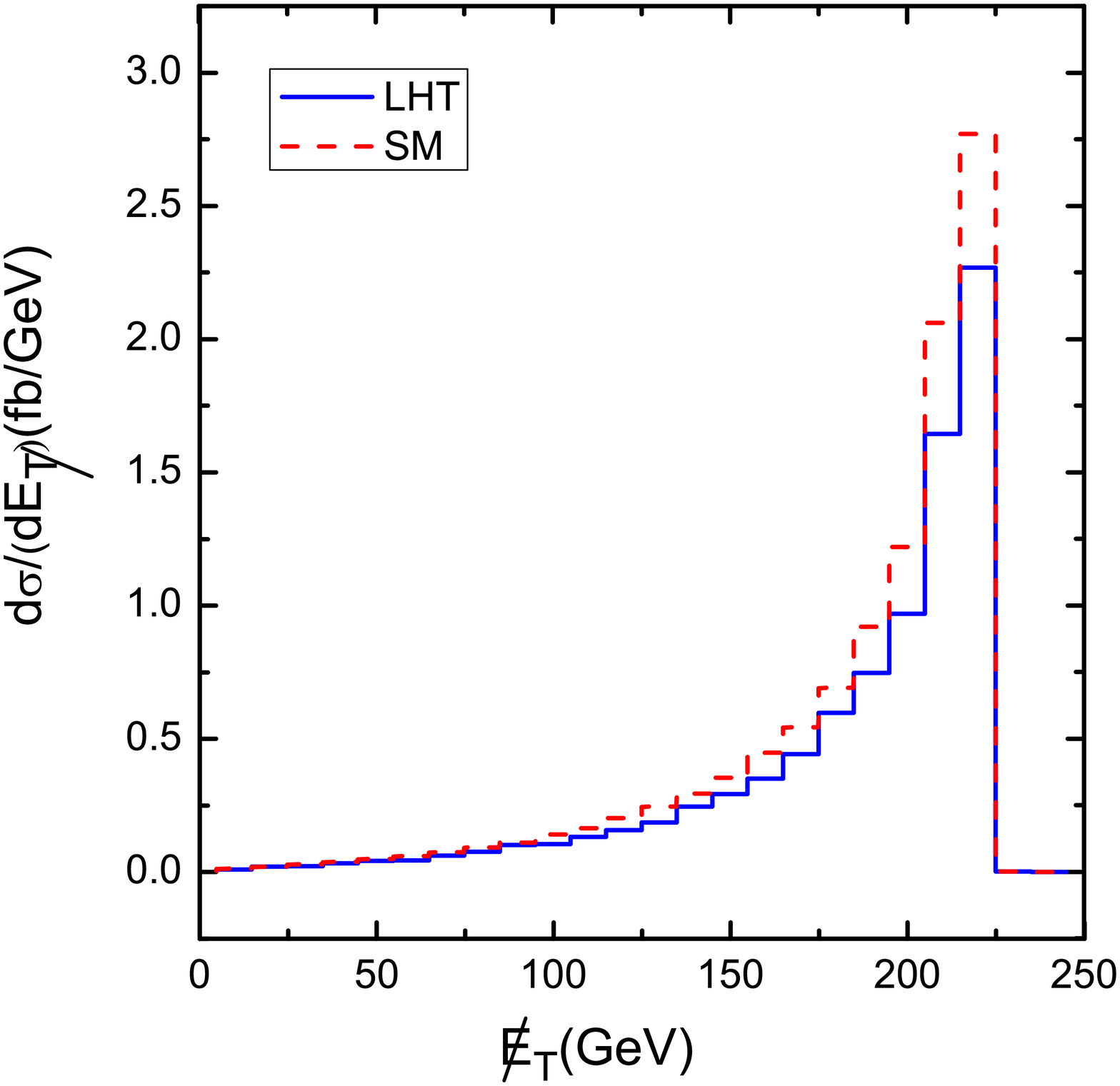}}\vspace{-0.8cm}
\scalebox{0.25}{\epsfig{file=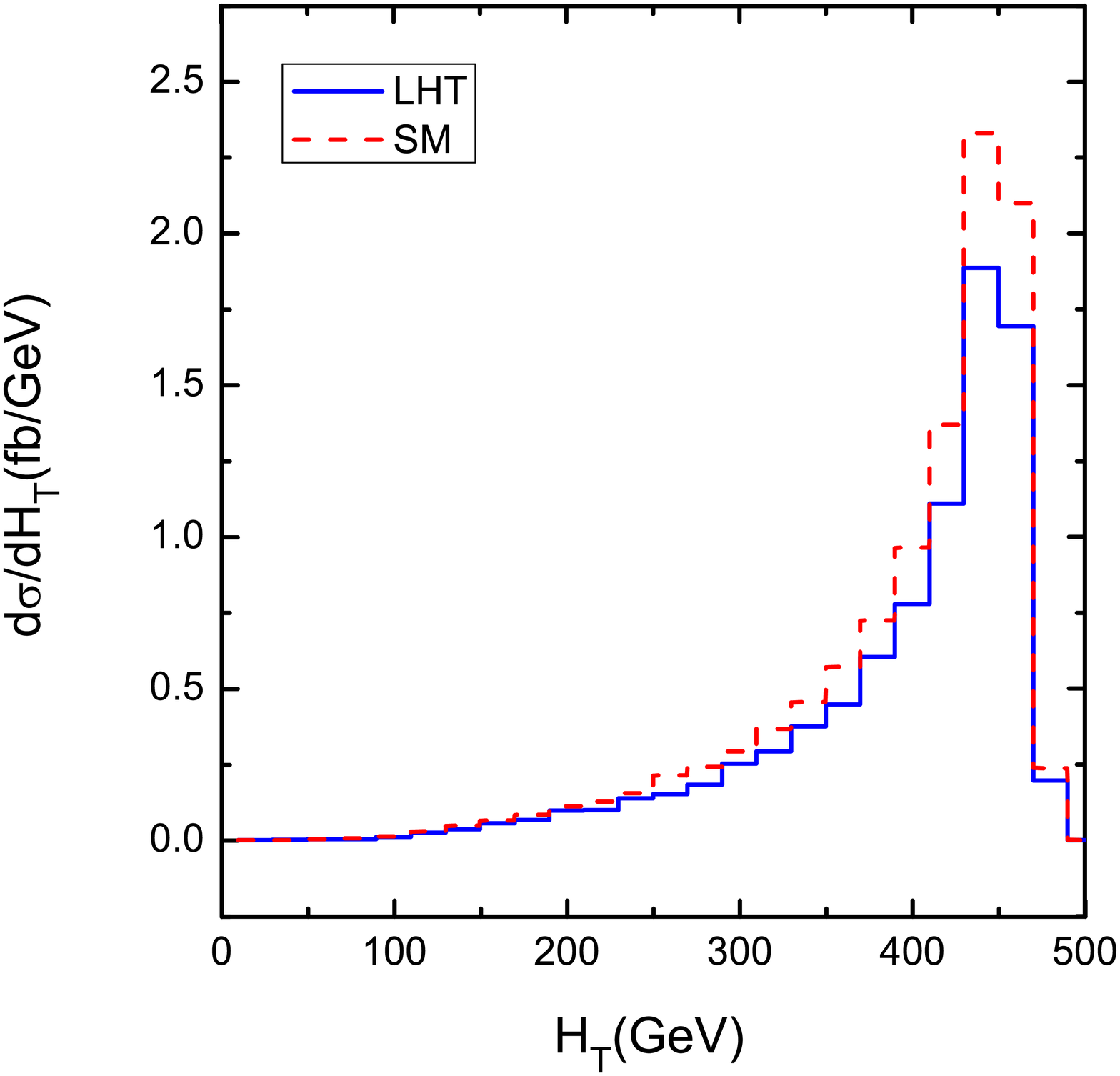}}\hspace{-0.8cm}
\scalebox{0.25}{\epsfig{file=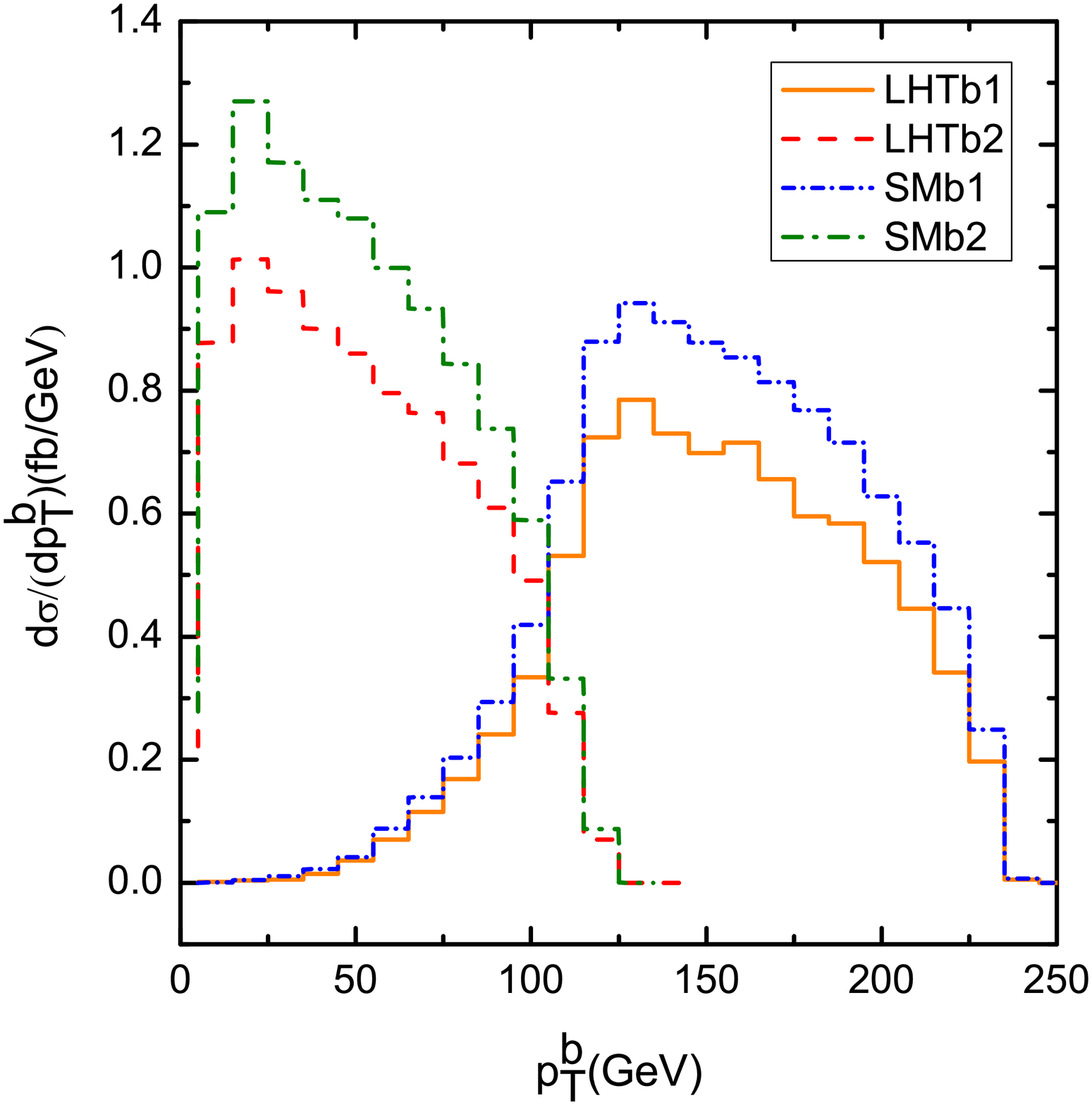}}\vspace{-0.8cm}\caption{$\Delta
R_{bb}$, $\met$, $H_{T}$ and $p^{b}_{T}$ distributions of
$e^{+}e^{-} \to ZH$ in the LHT and SM through the production of
$e^{+}e^{-}\rightarrow ZH\rightarrow \nu_{l}\bar{\nu_{l}}b\bar{b}$
for $\sqrt{s}=500$ GeV, $f=700$ GeV. } \label{observability}
\end{center}
\end{figure}

In order to provide more information of the single Higgs-boson
production, we display some kinematical distributions of final
states by using Madgraph5\cite{mad5}. In Fig.4, we show the
distributions of the production process $e^{+}e^{-}\rightarrow ZH$
for $\sqrt{s}=500$ GeV, $f=700$ GeV. We choose the
$e^{+}e^{-}\rightarrow ZH\rightarrow
\nu_{l}\bar{\nu_{l}}b\bar{b}$($l=e,\mu,\tau$) as the final states
and the relevant Feynman diagram is shown in Fig.3(a). We display
the separation between the two b-jets from Higgs boson ($\Delta
R_{bb} \equiv \sqrt{(\Delta \phi)^2 + (\Delta \eta)^2}$ ), the
missing energy $\met$, total transverse energy $H_{T}$ and the
transverse momentum $p^{b}_{T}$ of di b-tagged jets in the LHT and
the SM, respectively. We can see that the peak of the $\Delta
R_{bb}$ is at $\Delta R_{bb}\sim 1$, the peak of the missing energy
is at $\met\sim 220$ GeV, and the peak of the total transverse
energy is at $H_{T}\sim 440$ GeV. The transverse momentum
$p^{b}_{T}$ for the two b-jets is different, one peak is at
$p^{b1}_{T}\sim 130$ GeV and the other peak is at $p^{b2}_{T}\sim
20$ GeV.

\begin{figure}[htbp]
\begin{center}
\scalebox{0.25}{\epsfig{file=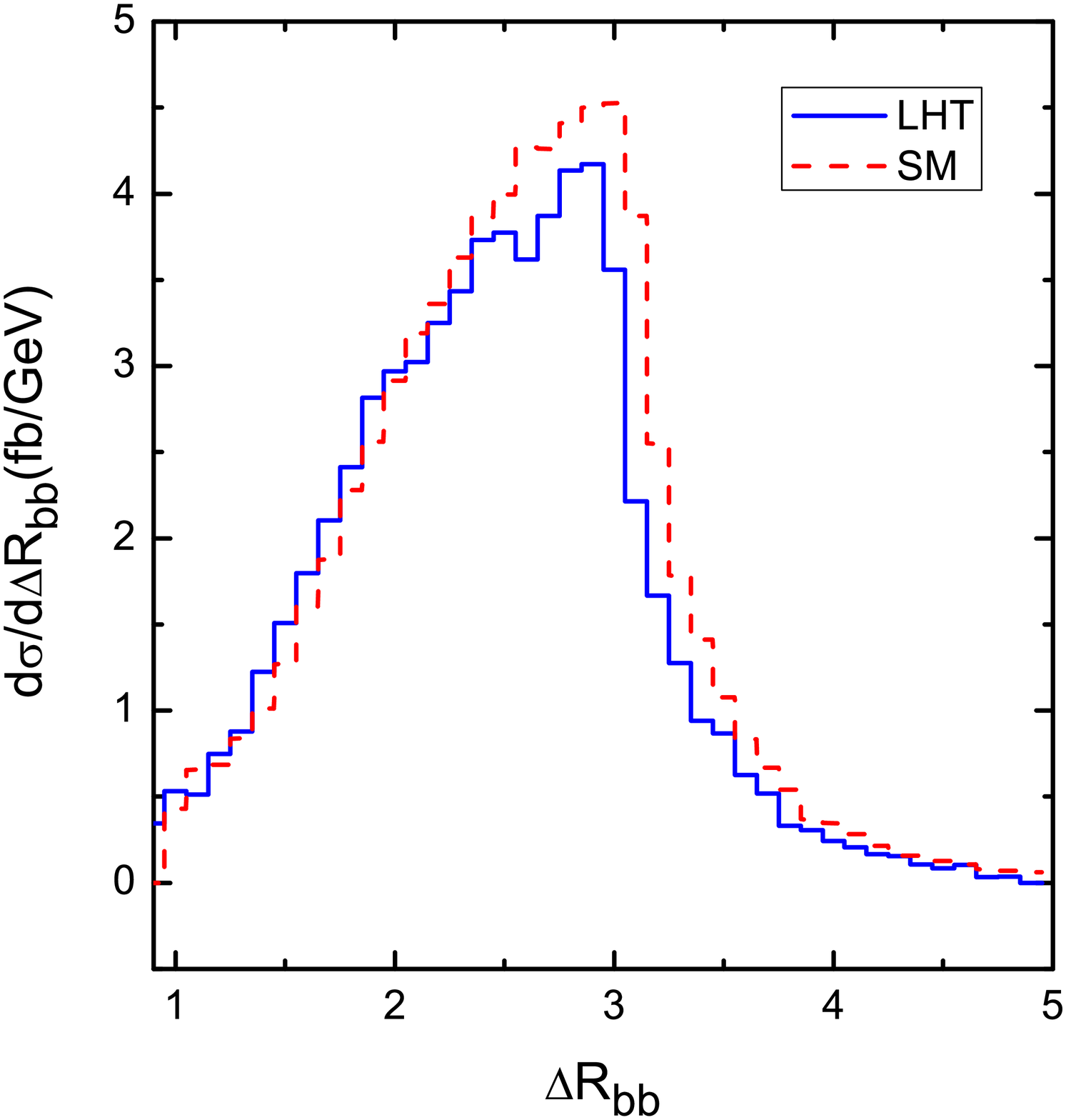}}\hspace{-0.8cm}
\scalebox{0.25}{\epsfig{file=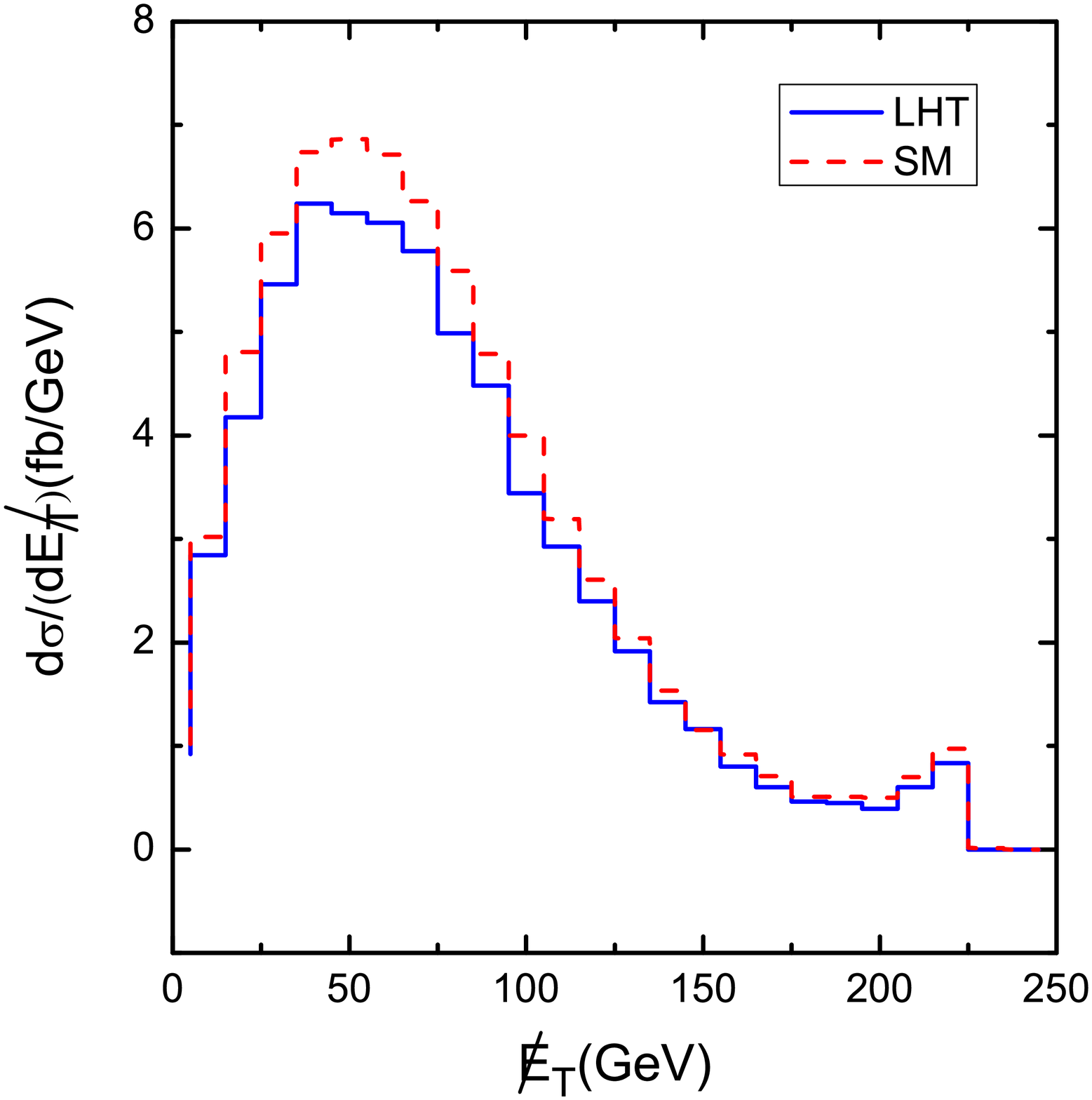}}\vspace{-0.8cm}
\scalebox{0.25}{\epsfig{file=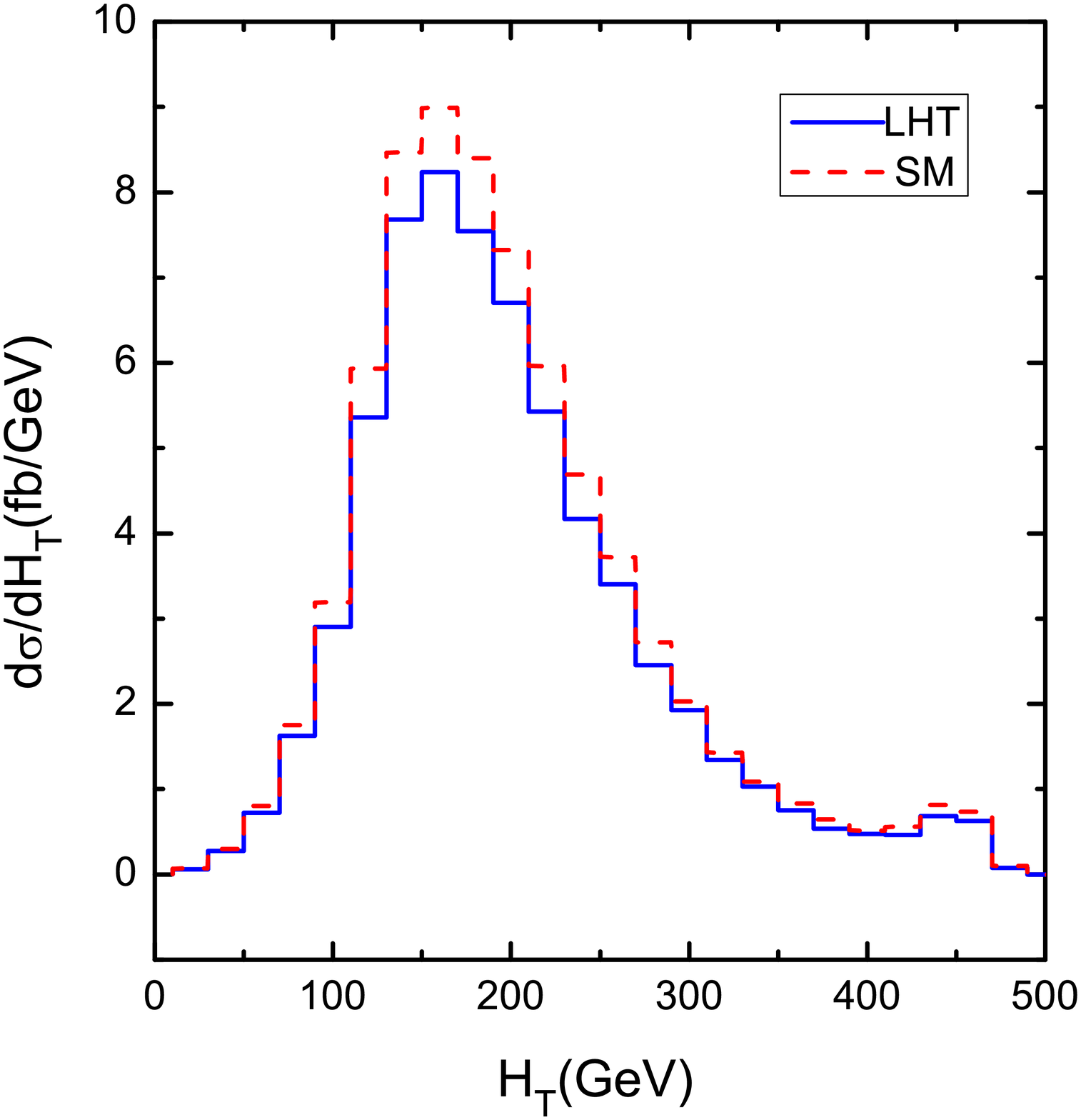}}\hspace{-0.8cm}
\scalebox{0.25}{\epsfig{file=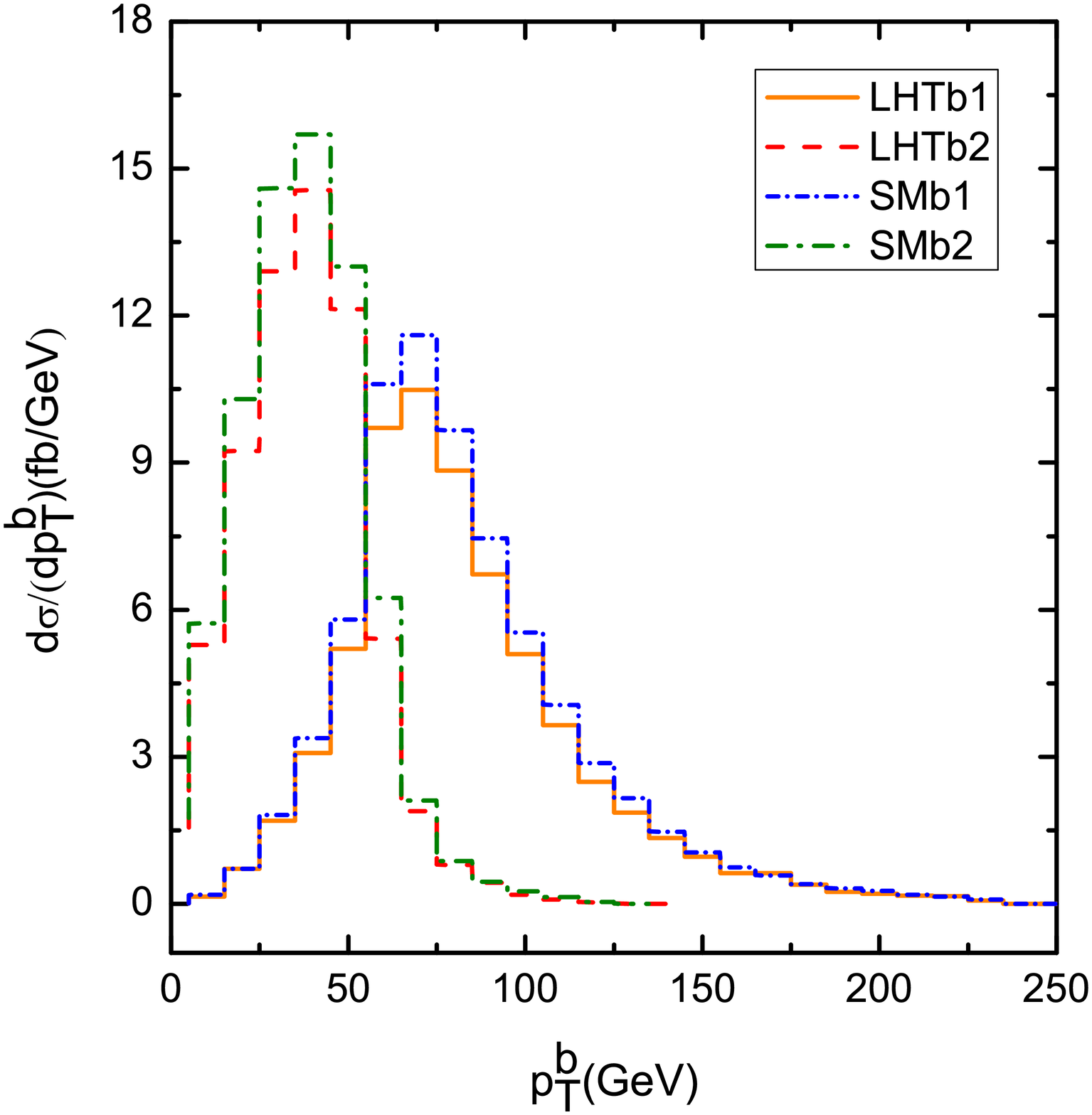}}\vspace{-0.8cm}\caption{$\Delta
R_{bb}$, $\met$, $H_{T}$ and $p^{b}_{T}$ distributions of
$e^{+}e^{-} \to \nu_{e}\bar{\nu_{e}}H$ in the LHT and SM through the
production of $e^{+}e^{-}\rightarrow
\nu_{e}\bar{\nu_{e}}H\rightarrow \nu_{e}\bar{\nu_{e}}b\bar{b}$ for
$\sqrt{s}=500$ GeV, $f=700$ GeV.} \label{observability}
\end{center}
\end{figure}

In Fig.5, we show the distributions of production process
$e^{+}e^{-} \to \nu_{e}\bar{\nu_{e}}H$ for $\sqrt{s}=500$ GeV,
$f=700$ GeV. We choose the $e^{+}e^{-}\rightarrow
\nu_{e}\bar{\nu_{e}}H\rightarrow \nu_{e}\bar{\nu_{e}}b\bar{b}$ as
the final states and the relevant Feynman diagrams are shown in
Fig.3(b). We display $\Delta R_{bb}$, $\not{\hspace*{-0.2cm}E}_T$,
$H_{T}$ and $p^{b}_{T}$ in the LHT and the SM, respectively. We can
see that the peak of the $\Delta R_{bb}$ is at $\Delta R_{bb}\sim
3$, which means that the two b-jets incline to fly back-to-back. The
peak of the missing energy is at $\met \sim 50$ GeV, and the peak of
the total transverse energy is at $H_{T}\sim 160$ GeV. The
transverse momentum $p^{b}_{T}$ for the two b-jets is different, one
peak is at $p^{b1}_{T}\sim 70$ GeV and the other peak is at
$p^{b2}_{T}\sim 40$ GeV.

From Fig.4 and Fig.5, we can see that the behaviour of the relevant
distributions in the LHT model is similar to that in the SM,  and
the LHT correction can obviously reduce the SM differential cross
section at around the peak.
\section{summary}

\noindent

In this paper, we studied the single Higgs-boson production at
$e^{+}e^{-}$ colliders in the LHT model. The main production
channels, such as $e^{+}e^{-}\rightarrow ZH$, $e^{+}e^{-}\rightarrow
\nu_{e}\bar{\nu_{e}}H$ and $e^{+}e^{-}\rightarrow e^{+}e^{-}H$, have
been taken into account. We calculated the production cross section
and the relative correction at the tree level. Considering the
latest constraints, we found that the relative correction of the
$ZH$ production channel can reach $-7.5\%$ and the relative
corrections of the $\nu_{e}\bar{\nu_{e}}H$ and $e^{+}e^{-}H$
production channels can both reach $-6.5\%$ for $\sqrt{s}=500$ GeV
with the lower limit of the scale $f$(=694 GeV), which is large
enough for people to detect the LHT effects at the future
$e^{+}e^{-}$ colliders. In order to investigate the observability,
some final state distributions of the production processes were
presented.

\section*{Acknowledgement}
This work is supported by the National Natural Science Foundation of
China under grant Nos.11347140, 11305049 and Specialized Research
Fund for the Doctoral Program of Higher Education under Grant
No.20134104120002.

\end{document}